\documentclass[12pt]{iopart}
\usepackage[dvips,dvipdfm]{graphicx}
\usepackage{amssymb}
\usepackage{fancybox}
\newcommand{\bea}{\begin{eqnarray}}
\newcommand{\eea}{\end{eqnarray}}

\newcommand{\bTheta}{\mbox{\boldmath{$\Theta$}}}

\newcommand{\bm}{\mbox{\boldmath{$m$}}}
\newcommand{\bh}{\mbox{\boldmath{$h$}}}

\newcommand{\bw}{\mbox{\boldmath{$w$}}}
\newcommand{\bx}{\mbox{\boldmath{$x$}}}

\newcommand{\be}{\mbox{\boldmath{$e$}}}

\newcommand{\by}{\mbox{\boldmath{$y$}}}

\newcommand{\bu}{\mbox{\boldmath{$u$}}}

\newcommand{\bzero}{\mbox{\boldmath{$0$}}}

\newcommand{\cL}{{\cal L}}
\newcommand{\cI}{{\cal I}}

\newcommand{\cQ}{{\cal Q}}

\newcommand{\mR}{\mathbb{R}}
\newcommand{\mN}{\mathbb{N}}
\newcommand{\qw}{q_w}
\newcommand{\qu}{q_u}
\newcommand{\hqw}{\widehat{q}_w}
\newcommand{\hqu}{\widehat{q}_u}
\newcommand{\cw}{\chi_w}
\newcommand{\cu}{\chi_u}
\newcommand{\hcw}{\widehat{\chi}_w}
\newcommand{\hcu}{\widehat{\chi}_u}
\newcommand{\bDelta}{\mbox{\boldmath{$\Delta$}}}
\newcommand{\cA}{{\cal A}}

\newcommand{\cqw}{{\cal Q}_w}
\newcommand{\cqu}{{\cal Q}_u}
\newcommand{\img}{{\rm i}}
\newcommand{\subst}{{\leftarrow}}

\begin{document}

\title[Classification of correlated patterns by perceptrons]{Perceptron 
capacity revisited: classification ability for correlated patterns}

\author{Takashi Shinzato and Yoshiyuki Kabashima}

\address{Department of Computational Intelligence and Systems Science, Tokyo
Institute of Technology, Yokohama 226-8502, Japan}
\ead{shinzato@sp.dis.titech.ac.jp, kaba@dis.titech.ac.jp}
\begin{abstract}
In this paper, we address the problem of how many randomly labeled patterns can be 
correctly classified by a single-layer perceptron when 
the patterns are correlated with each other. 
In order to solve this problem, two analytical schemes are 
developed based on the replica method and 
Thouless-Anderson-Palmer (TAP) approach by utilizing 
an integral formula concerning random rectangular matrices.
The validity and relevance of the developed 
methodologies are shown for one known result and two example problems. 
A message-passing algorithm to perform the TAP 
scheme is also presented. 
\end{abstract}

\maketitle

\section{Introduction}
Learning from examples is one of the most 
significant problems in information science, and (single-layer) perceptrons 
are often included in widely used devices for solving this problem. 
In the last two decades, the structural similarity between 
the learning problem and the statistical mechanics 
of disordered systems has been observed, thus promoting
cross-disciplinary research on perceptron learning with the use of 
methods from statistical mechanics 
\cite{LevinTishbySolla1990,GyorgyiTishby1990}. 
This research activity has successfully contributed
to the finding of various behaviors in the learning process of 
perceptrons \cite{WatkinRauBiehl1992,Engel2001,Nishimori2001}
and to the development of computationally 
feasible approximate learning 
algorithms \cite{OpperSaad2001,MezardParisiZecchina2002}
that had never been discovered by conventional 
approaches in information science, particularly for the 
non-asymptotic regimes in which the ratio between the numbers 
of examples $p$ and weight parameters $N$, $\alpha=p/N$, is $O(1)$. 

Although such statistical mechanical methodologies 
have been successfully applied to learning problems, 
there still remain several research directions to explore. 
Learning from correlated patterns is a typical example of such a problem. 
In most of the earlier studies, it was assumed,
for simplicity, that the input patterns used 
for learning were independently and identically distributed (IID)
\cite{WatkinRauBiehl1992,Engel2001,Nishimori2001}. 
However, this assumption is obviously not practical since 
real-world data is usually somewhat biased and correlated
across components, which makes it difficult to utilize the 
developed schemes directly for learning beyond a conceptual level. 
In order to increase the practical relevance of the 
statistical mechanical approach, it is necessary 
to generalize the approach to handle correlated patterns. 

As a first step for such a research direction, 
we address the problem of correctly classifying many randomly labeled patterns 
by a single-layer perceptron when 
the patterns are correlated with each other. 
In data analysis, problems of this kind are of practical 
importance as an assessment of null hypotheses that state no regularity 
represented by the perceptron underlies a given data set. 
In addition, recent deepening of the relations across learning, 
information and communication shows that the
perceptron can be utilized as a useful building block
for various coding schemes 
\cite{Tanaka2002,HosakaKabashimaNishimori2002,KanterKinzel2002,
MimuraOkada2006}. Therefore, exploration to handle learning 
from correlated patterns
may lead to the development of better schemes used 
for information and communication engineering. 

This paper is organized as follows. 
In the next section, we introduce the problem we are studying. In section 3, 
which is the main part of
this article, we develop two schemes for analyzing the 
problem on the basis of the replica method and Thouless-Anderson-Palmer 
(TAP) approach. Statistical mechanical techniques
that can handle correlated patterns have already been 
developed by Opper and Winther \cite{OpperWinther2001L,
OpperWinther2001, OpperWinther2005}. However, their schemes, 
which apply to densely connected networks of two-body interactions, 
are highly general, and therefore properties that hold specifically 
for perceptrons are not fully utilized. Hence, in this paper, we offer 
specific methodologies that can be utilized for perceptron type networks. 
We show that an integral formula provided for ensembles
of rectangular random matrices plays important roles for 
the provided methods. A message-passing algorithm to 
solve the developed TAP scheme is also presented. 
In section 4, the validity and utility of the methods
are shown by applications to one known result and two example problems. 
The final section is a summary.

\section{Problem definition}
In a general scenario, for an $N$-dimensional input pattern vector
$\bx$, a perceptron which is parameterized by an $N$-dimensional 
weight vector $\bw$ can be identified with an indicator function 
of class label $y=\pm 1$, 
\begin{eqnarray}
\cI \left (y|\Delta \right ), 
\label{perceptron}
\end{eqnarray}
where $\cI (y|\Delta)=1-\cI (-y|\Delta)$ takes $1$ or $0$ depending on 
the value of internal potential $ \Delta = N^{-1/2} \bw \cdot \bx$. 
Prefactor $N^{-1/2}$ is introduced to keep relevant variables
$O(1)$ as $N \to \infty$. 
Equation (\ref{perceptron}) indicates that a perceptron 
specified by $\bw$ correctly classifies a given labeled pattern $(\bx,y)$ if 
$\cI \left (y|\Delta \right )=1$;
otherwise, it does not make the correct classification. Let us suppose that a set of 
patterns $\bx_1,\bx_2,\ldots,\bx_p$ is given. 
The problem we consider here is whether the perceptron 
can typically classify the patterns correctly by only adjusting 
$\bw$ when the class label of each pattern 
$\bx_\mu$, $y_\mu \in \{+1,-1\}$, is 
independently and randomly assigned with 
a probability of 1/2 for $\mu=1,2,\ldots,p$
as $N$ and $p$ tend to infinity, keeping the
pattern ratio 
$\alpha=p/N $ of the order of unity. 

In general, entries of pattern matrix $X=N^{-1/2}(\bx_1,\bx_2,
\ldots,\bx_p)^{\rm T}$ are correlated with each 
other, where ${\rm T}$ denotes the matrix transpose. 
As a basis for dealing with such correlations, 
we introduce an expression of the singular value decomposition 
\begin{eqnarray}
X=U^{\rm T} D V, 
\label{SVD}
\end{eqnarray}
of the pattern matrix $X$, where $D={\rm diag}(d_k)$ is a $p \times N$
diagonal matrix composed of singular values
$d_k$ $(k=1,2,\ldots, {\rm min}(p,N))$, and 
$U$ and $V$ are $p \times p$ and $N \times N$ orthogonal matrices, 
respectively.
${\rm min}(p,N)$ denotes the 
lesser value of $p$ and $N$. 
Linear algebra guarantees that an arbitrary 
$p \times N$ matrix can be decomposed according to 
equation (\ref{SVD}). 
The singular values $d_k$ are linked to eigenvalues 
of the correlation matrix $X^{\rm T} X$, 
$\lambda_k$ $(k=1,2,\ldots,N)$, as
$\lambda_k=d_k^2$ $(k=1,2,\ldots, {\rm min}(p,N))$ and 
$0$ otherwise. 
The orthogonal matrices $U$ and $V$ constitute
the eigen bases of correlation matrices
$XX^{\rm T}$ and $X^{\rm T}X$, respectively. 
In order to handle correlations in $X$ analytically, we 
assume that $U$ and $V$ are uniformly and independently 
generated from the Haar measures of 
$p \times p$ and $N \times N$ orthogonal matrices, respectively, 
and that the empirical eigenvalue spectrum  of $X^{\rm T}X$,
$N^{-1}\sum_{k=1}^N \delta(\lambda-\lambda_k)=
(1-{\rm min}(p,N)/N) \delta(\lambda)+ 
N^{-1} \sum_{k=1}^{{\rm min}(p,N)} \delta(\lambda-d_k^2)$, 
converges to a certain specific distribution $\rho(\lambda)$
in the large system limit of $N,p \to \infty$, $\alpha =p/N \sim O(1)$. 
Controlling $\rho(\lambda)$ allows us to characterize various
second-order correlations in pattern matrix $X$. 

For generality and analytical tractability, let us 
assume that $\bw$ obeys a factorizable distribution 
$P(\bw)=\prod_{i=1}^N P(w_i)$ {\em a priori}.
Given a labeled pattern set $\xi^p=(X,\by)$, where
$\by=(y_1,y_2,\ldots,y_p)$, it is possible to assess 
the volumes of $\bw$ that are compatible with $\xi^p$ as
\begin{eqnarray}
V(\xi^p)=\mathop{\rm Tr}_{\bw }
\prod_{i=1}^N P(w_i) \prod_{\mu=1}^p \cI(y_\mu|\Delta_\mu), 
\label{gardner}
\end{eqnarray}
where $\Delta_\mu=N^{-1/2} \bw \cdot \bx_\mu $ $(\mu=1,2,\ldots,p)$ 
and ${\rm Tr}_{\bw}$ denotes the summation (or integral) over
all possible states of $\bw$. 
Equation (\ref{gardner}), which is sometimes referred 
to as the {\em Gardner volume}, is used for assessing whether 
$\xi^p$ can be classified by a given type of perceptron
because it is possible to choose an appropriate $\bw$ that 
is fully consistent with $\xi^p$ if and only 
if $V(\xi^p)$ does not vanish \cite{Gardner1988}.

In the large system limit, $V(\xi^p)$ typically 
vanishes and, therefore, $\xi^p$ cannot be correctly 
classified by perceptrons of the given type when $\alpha$ becomes 
larger than a certain critical value $\alpha_c$, 
which is often termed {\em perceptron capacity} \cite{Cover1965,Gardner1988}. 
Since the mid-1980s, 
much effort has been made in the cross-disciplinary 
field of statistical mechanics and information 
science to assess $\alpha_c$ in various systems \cite{Engel2001}: 
in particular, for pattern matrices 
entries of which are independently drawn from an identical 
distribution of zero mean 
and variance $N^{-1}$.
%entries which have an IID  Variance $N^{-1}$
Such situations are characterized by the Mar\u{c}enko-Pastur 
law $\rho(\lambda)=[1-\alpha]^+\delta(\lambda)+
(2 \pi)^{-1}\lambda^{-1}
\sqrt{[\lambda-\lambda_{-}]^+ [\lambda_{+}-\lambda]^+}$
in the current framework, where $[x]^+=x$ for $x>0$ and $0$, otherwise, 
and $\lambda_{\pm}=\left (\sqrt{\alpha}\pm 1 \right )^2$ 
\cite{TulinoVerdu2004}. 
However, it seems that little is known 
about how the correlations in pattern matrices, 
which are characterized by $\rho(\lambda)$ here, 
influence the perceptron capacity $\alpha_c$. 
Therefore, the main objective of the present article
is to answer this question.

\section{Analysis}
\subsection{A generalization of the Itzykson-Zuber integral}
The expression
\begin{eqnarray}
V(\xi^p)&=&\mathop{\rm Tr}_{\bw}\prod_{i=1}^N P(w_i)\prod_{\mu=1}^p
\left (
\int d \Delta_\mu \cI(y_\mu|\Delta_\mu) 
\delta(\Delta_\mu-N^{-1/2} \bw \cdot \bx_\mu) \right )\cr
&=&\int \prod_{\mu=1}^p 
\left (\frac{d u_\mu d \Delta_\mu}{2 \pi} 
\exp \left [-\img u_\mu \Delta_\mu \right ] \cI(y_\mu|\Delta_\mu) \right )
\mathop{\rm Tr}_{\bw}P(w_i) \exp \left [\img \bu^{\rm T}X \bw \right ] \cr
&=& \mathop{\rm Tr}_{\bu,\bw} \prod_{\mu=1}^p \widehat{\cI}_{y_\mu}(u_\mu)
\prod_{i=1}^N P(w_i) \exp \left [ \img \bu^{\rm T}X \bw \right ] 
\label{gardner2}
\end{eqnarray}
constitutes the basis for analyzing the behavior of equation (\ref{gardner}), 
where $\img=\sqrt{-1}$, $\bu=(u_1,u_2,\ldots,u_p)^{\rm T}$ and 
$\widehat{\cI}_{y_\mu}(u_\mu)=\int d \Delta_\mu 
\exp \left [-\img u_\mu \Delta_\mu \right ]\cI(y_\mu|\Delta_\mu)/(2 \pi)$. 
In order to evaluate the average of 
$V(\xi^p)$, we substitute 
equation (\ref{SVD}) into equation (\ref{gardner2}) and take the 
average with respect to the orthogonal matrices 
$U$ and $V$. For this assessment, it is worthwhile to note that 
for the fixed sets of dynamical variables 
$\bw$ and $\bu$, $\widetilde{\bw}=V \bw$ and $\widetilde{\bu}=U\bu$
behave as continuous random variables that are uniformly 
generated under the strict constraints
\begin{eqnarray}
&& \frac{1}{N}|\widetilde{\bw}|^2 =\frac{1}{N} |\bw|^2 = Q_w, \label{w_norm} \\
&& \frac{1}{p}|\widetilde{\bu}|^2 =\frac{1}{p} |\bu|^2 = Q_u, \label{u_norm} 
\end{eqnarray}
when $U$ and $V$ are independently and 
uniformly generated from the Haar measures. 
In the limit as $N,p \to \infty$, keeping $\alpha=p/N \sim O(1)$, 
this yields the expression 
\begin{eqnarray}
&&\frac{1}{N} \ln \left [\overline{\exp \left [{\rm i} \bu^{\rm T}X \bw  
\right ]} \right ] \cr
&& =\frac{1}{N} \ln \left [ 
\frac{
\int d \widetilde{\bw} d \widetilde{\bu}
\delta \left ( |\widetilde{\bw}|^2- N Q_w\right )
\delta \left ( |\widetilde{\bu}|^2- p Q_u\right )
\exp \left [{\rm i} \widetilde{\bu}^{\rm T}
D
\widetilde{\bw  }
\right ]}
{\int d \widetilde{\bw} d \widetilde{\bu}
\delta \left ( |\widetilde{\bw}|^2- N Q_w\right )
\delta \left ( |\widetilde{\bu}|^2- p Q_u\right )}
\right ] \cr
&&=F(Q_w,Q_u), 
\label{generative_F}
\end{eqnarray}
where $\overline{\cdots}$ denotes 
averaging with respect to the Haar measures, 
the function $F(x,y)$ is assessed as
\begin{eqnarray}
&&F(x,y)=\mathop{\rm Extr}_{\Lambda_x,\Lambda_y}
\left \{-
\frac{1}{2}\left \langle \ln (\Lambda_x \Lambda_y+\lambda)
\right \rangle_\rho 
-\frac{\alpha-1}{2}\ln \Lambda_y+\frac{\Lambda_x x}{2}
+\frac{\alpha \Lambda_y y}{2}\right \} \cr
&&\phantom{F(x,y)=}
-\frac{1}{2}\ln x -\frac{\alpha}{2}\ln y-\frac{1+\alpha}{2}, 
\label{F_func}
\end{eqnarray}
and $\left \langle \cdots \right \rangle_\rho$ indicates 
averaging with respect to the asymptotic eigenvalue spectrum of 
$X^{\rm T}X$, $\rho(\lambda)$ \cite{Kabashima2007}. 
The derivation of equations (\ref{generative_F}) and (\ref{F_func}) 
is shown in Appendix A. 
$\mathop{\rm Extr}_{\theta} \left \{ \cdots \right \}$ 
represents extremization with respect to $\theta$. 
%%%%
This corresponds to the saddle point assessment of a
complex integral and does not necessarily 
mean the operation of a minimum or maximum. Expressions analogous to 
these equations %(\ref{generative_F}) and (\ref{F_func}) 
are known as the Itzykson-Zuber integral or $G$-function 
for ensembles of square (symmetric) matrices 
\cite{ItzyksonZuber1980,VoiculescuDykemaNica1992,
MarinariParisiRitort1994,ParisiPotters1995,
CherrierDeanLefevre2003,TakedaUdaKabashima2006,
TakedaHatabuKabashima2007,MullerGuoMoustakas2007,Tanaka2007}. 
%%%%
Equation (\ref{generative_F}) implies that the annealed average of 
equation (\ref{gardner}) is evaluated as 
\begin{eqnarray}
\frac{1}{N}\ln 
\left [V(\xi^p)  \right ]_{\xi^p} =\mathop{\rm Extr}_{Q_w,Q_u}
\left \{F(Q_w,Q_u)+A_w(Q_w)+\alpha A_u(Q_u) \right \}, 
\label{anneal_action}
\end{eqnarray}
where $\left [\cdots \right ]_{\xi^p}=2^{-p} \mathop{\rm Tr}_{\by}
\overline{(\cdots)}$ represents the average with respect to 
a set of randomly labeled patterns $\xi^p$ and 
\begin{eqnarray}
A_w(Q_w)=\mathop{\rm Extr}_{\widehat{Q}_w}
\left
\{ \frac{\widehat{Q}_w Q_w}{2} + 
\ln \left [\mathop{\rm Tr}_{w} P(w)\exp \left [-\frac{\widehat{Q}_w}{2}w^2 \right ]
\right ]
\right \}, 
\label{AS}
\end{eqnarray}
\begin{eqnarray}
A_u(Q_u)=\mathop{\rm Extr}_{\widehat{Q}_u}
\left
\{ \frac{\widehat{Q}_u Q_u}{2} + 
\ln \left [\frac{1}{2}\mathop{\rm Tr}_{u,y} 
\widehat{\cI}_y(u)\exp \left [-\frac{\widehat{Q}_u}{2}u^2 \right ]
\right ]
\right \}. 
\label{Au}
\end{eqnarray}
Normalization constraints $\mathop{\rm Tr}_y \cI(y|\Delta)=1$ 
guarantee that $\left [V(\xi^p) \right ]_{\xi^p}=2^{-p}$, 
which implies that for any $\bw$ 
the probability that each randomly labeled 
pattern $(\bx_\mu,y_\mu)$ ($\mu=1,2,\ldots,p$) is correctly 
classified is equally $1/2$ and, therefore, 
the size of feasible volume $V(\xi^p)$ 
decreases as $2^{-p}$ on average,
regardless of correlations in $X$.  
In addition, in conjunction with equations (\ref{anneal_action}), 
(\ref{AS}) and (\ref{Au}), this
implies that $Q_w={\rm Tr}_w w^2 P(w)$, $Q_u=0$, 
$\widehat{Q}_w=0$ and $\widehat{Q}_u= 
\alpha^{-1} Q_w \left \langle \lambda \right \rangle_\rho$.  
The physical implication is that, due to the central limit theorem, 
$\bDelta=(\Delta_1,\Delta_2,\ldots,\Delta_p)^{\rm T}$
follows an isotropic Gaussian distribution
\begin{eqnarray}
P(\bDelta)=\frac{1}{(2 \pi \widehat{Q}_u)^{p/2}} 
\exp \left [-\frac{|\bDelta|^2}{2\widehat{Q}_u  } \right ]
= \frac{\alpha^{p/2}}{(2 \pi Q_w \left \langle \lambda \right \rangle_\rho
)^{p/2}}
\exp \left [
-\frac{\alpha |\bDelta|^2}{2 Q_w \left \langle \lambda \right \rangle_\rho}
\right ], 
\label{delta_dist} 
\end{eqnarray}
in the limit as $N,p \to \infty$, $\alpha=p/N \sim O(1)$ 
when $\bw$ is generated from $P(\bw)=\prod_{i=1}^N P(w_i)$, and 
$U$ and $V$ are independently and uniformly generated from the Haar measures. 

\subsection{Replica analysis}
Now we are ready to analyze the typical behavior of equation (\ref{gardner}). 
Because $\xi^p$ is a set of quenched random variables, we resort to 
the replica method \cite{SherringtonKirkpatrick1975,MPV1987,Dotzenko2001}. 
This indicates that we evaluate the $n$-th moment of 
$V(\xi^p)$ for natural numbers $n\in \mN$ as
\begin{eqnarray}
\left [ V^n (\xi^p) 
\right ]_{\xi^p}
%%&=&2^{-p} \mathop{\rm Tr}_{\by}\overline{ V^n(\xi^p)}\cr
&=&\mathop{\rm Tr}_{
\{\bu^a\},\{\bw^a\}}
\prod_{\mu=1}^p
\left (\frac{1}{2} \mathop{\rm Tr}_{y_\mu}
\prod_{a=1}^n \widehat{\cI}_{y_\mu}(u^{a}_\mu ) \right )
\times \prod_{i=1}^N 
\left (\prod_{a=1}^n P(w_i^a) \right ) \cr
&& \phantom{\left [ Z_P^n (\xi^p) 
\right ]_{\xi^p}=}
\times \overline{\exp \left [{\rm i}\sum_{a=1}^n 
(\bu^a)^{\rm T}X \bw^a \right ]}, 
\label{moments_natural}
\end{eqnarray}
and assess the quenched average of free energy with respect to 
the labeled pattern set $\xi^p$ as 
$ N^{-1}\left [ \ln  V (\xi^p) \right ]_{\xi^p} =\lim_{n \to 0}
\frac{\partial}{\partial n}
N^{-1}
\ln \left [V^n (\xi^p) \right ]_{\xi^p}$
by analytically continuing expressions obtained for
equation (\ref{moments_natural}) from 
$n \in \mN$ to real numbers $n \in \mR$. 
Here, $\{\bw^a\}$ and $\{\bu^a\}$ represent sets of dynamical 
variables $\bw^1,\ldots,\bw^n$ and
$\bu^1,\ldots,\bu^n$, respectively, 
where $1,2,\ldots,n$ denote the $n$ replicas of perceptrons.

For this procedure, an explanation similar to that for the evaluation of 
equation (\ref{generative_F}) is useful. Namely, for fixed sets of 
dynamical variables $\{\bu^a\}$ and $\{\bw^a\}$, 
$\widetilde{\bu}^a=U \bu^a$ and $\widetilde{\bw}^a=V \bw^a$ 
behave as continuous random variables which satisfy strict constraints
\begin{eqnarray}
&&\frac{1}{N} \widetilde{\bw}^a \cdot \widetilde{\bw}^b
=\frac{1}{N} {\bw}^a \cdot {\bw}^b =\qw^{ab},  \label{qsab}\\
&&\frac{1}{p} \widetilde{\bu}^a \cdot \widetilde{\bu}^b
=\frac{1}{p} {\bu}^a \cdot {\bu}^b =\qu^{ab}, \label{quab}
\end{eqnarray}
$(a,b=1,\ldots,n)$
when $U$ and $V$ are 
independently and uniformly generated from the Haar measures. 
This indicates that equation (\ref{moments_natural}) can be 
evaluated by the saddle point method with respect to 
sets of macroscopic parameters $\cqw=(\qw^{ab})$ and 
$\cqu=(\qu^{ab})$ 
in the limit as $N,p \to \infty$, $\alpha=p/N \sim O(1)$. 
In addition, intrinsic permutation symmetry among replicas indicates
that it is natural to assume that 
$n \times n$ matrices $\cqw$ and $\cqu$ are of the replica symmetric 
(RS) form 
\begin{eqnarray}
\cqw&=&\left (
\begin{array}{cccc}
\cw+q_w &q_w    & \ldots & q_w \cr
q_w    &\cw+q_w & \ldots & q_w \cr
\vdots  &\vdots  & \ddots & \vdots \cr
q_w    &q_w    & \ldots & \cw +q_w 
\end{array}
\right )\cr
&=& E \times \left (
\begin{array}{cccccc}
\cw+n q_w & \vline & 0 & 0 & \ldots & 0 \cr
\hline
0 & \vline &\cw & 0& \ldots & 0 \cr
0 & \vline & 0  & \cw & \ldots & 0 \cr
\vdots & \vline & \vdots & \vdots & \ddots &\vdots\cr
0 & \vline & 0 & 0 & \ldots & \cw
\end{array}
\right ) \times E^{\rm T}, 
\label{Qs}
\end{eqnarray}
and 
\begin{eqnarray}
\cqu&=&\left (
\begin{array}{cccc}
\cu-q_u &-q_u    & \ldots & -q_u \cr
-q_u    &\cu-q_u & \ldots & -q_u \cr
\vdots  &\vdots  & \ddots & \vdots \cr
-q_u    &-q_u    & \ldots & \cu -q_u 
\end{array}
\right ) \cr
&=&E \times \left (
\begin{array}{cccccc}
\cu-nq_u      &   \vline  & 0 & 0 & \ldots & 0 \cr
\hline 
0             &   \vline  & \cu & 0& \ldots & 0 \cr
0             &   \vline  &  0  & \cu & \ldots & 0 \cr
\vdots        &   \vline  & \vdots & \vdots & \ddots &\vdots\cr
0        &   \vline  & 0 & 0 & \ldots & \cu
\end{array}
\right ) \times E^{\rm T}, 
\label{Qu}
\end{eqnarray}
at the saddle point. Here, 
$E=(\be_1,\be_2,\ldots,\be_n)$ denotes an $n$-dimensional
orthonormal basis composed of 
$\be_1=(n^{-1/2},n^{-1/2},\ldots,n^{-1/2})^{\rm T}$ 
and $n-1$ orthonormal vectors $\be_2, \be_3,\ldots, \be_n$, 
which are orthogonal to $\be_1$. 
Equations (\ref{Qs}) and (\ref{Qu}) indicate that 
under the RS ansatz, the $n$ replicas that are coupled with each other
in equations (\ref{qsab}) and (\ref{quab})
can be decoupled by rotating $\{\widetilde{\bw}^a\}$ and 
$\{\widetilde{\bu}^a\}$ with respect to the replica coordinates 
simultaneously with the use of the identical orthogonal matrix $E$. 
The already decoupled expression $\sum_{a=1}^n 
({\bu}^a)^{\rm T}X {\bw}^a
=\sum_{a=1}^n 
(\widetilde{\bu}^a)^{\rm T}D \widetilde{\bw}^a$
is kept invariant under this rotation.
%because $\{\widetilde{\bw}^a\}$ and $\{\widetilde{\bu}^a\}$ are 
%simultaneously transformed using the 
%identical orthogonal matrix $E$. 
These operations imply that, in the new coordinates, the average 
with respect to
$U$ and $V$ over uniform distributions of the Haar measures 
can be evaluated individually
for each of the $n$ decoupled modes, 
which yields
\begin{eqnarray}
&& \frac{1}{N} \ln \left [ 
\overline{\exp \left [
\img \sum_{a=1}^n ({\bu}^a)^{\rm T}X {\bw}^a
 \right ] }\right ] 
\cr
&&= 
\frac{1}{N} \ln \left [ 
\frac{\int \prod_{a=1}^n d\widetilde{\bw}^a d\widetilde{\bu}^a 
{\cal C}_{\rm coupled}
\exp \left [
\img \sum_{a=1}^n (\widetilde{\bu}^a)^{\rm T}D \widetilde{\bw}^a
 \right ] 
}{\int \prod_{a=1}^n d\widetilde{\bw}^a d\widetilde{\bu}^a 
{\cal C}_{\rm coupled}
} \right ]\cr
&&= 
\frac{1}{N} \ln \left [ 
\frac{\int \prod_{a=1}^n d\widetilde{\bw}^a d\widetilde{\bu}^a 
{\cal C}_{\rm decoupled}
\exp \left [
\img \sum_{a=1}^n (\widetilde{\bu}^a)^{\rm T}D \widetilde{\bw}^a
 \right ] 
}{\int \prod_{a=1}^n d\widetilde{\bw}^a d\widetilde{\bu}^a 
{\cal C}_{\rm decoupled}
} \right ]\cr
&& =F(\cw+n q_w,\cu-nq_u)+ (n-1) F(\cw,\cu), 
\label{nthmoments}
\end{eqnarray}
where 
\begin{eqnarray}
{\cal C}_{\rm coupled}
=\prod_{a=1}^n \delta (|\widetilde{\bw}^a|^2 -
N(\cw+ q_w))
\prod_{a>b}\delta (\widetilde{\bw}^a
\cdot\widetilde{\bw}^b  
-Nq_w) \cr
\phantom{{\cal C}_{\rm coupled}=} \times
\prod_{a=1}^n \delta (|\widetilde{\bu}^a|^2 -
p(\cu- q_u))
\prod_{a>b}\delta (\widetilde{\bu}^a
\cdot\widetilde{\bu}^b  +pq_u), 
\label{const_w}
\end{eqnarray}
and
\begin{eqnarray}
{\cal C}_{\rm decoupled}=
\delta(|\widetilde{\bw}^1|^2-N(\cw+nq_w))\prod_{a=2}^n 
\delta(|\widetilde{\bw}^a|^2-Nq_w) \cr
\phantom{{\cal C}_{\rm decoupled}=}
\times \delta(|\widetilde{\bu}^1|^2-p(\cu-nq_u))\prod_{a=2}^n 
\delta(|\widetilde{\bu}^a|^2+pq_u). 
\label{const_u}
\end{eqnarray}
Equation (\ref{nthmoments})
and evaluation of the volumes of dynamical variables 
$\{\bw^a\}$ and $\{\bu^a\}$ under constraints (\ref{qsab}) and (\ref{quab}) 
of the RS ansatz (\ref{Qs}) and (\ref{Qu}) 
provide an expression for the average free energy
\begin{eqnarray}
&& \frac{1}{N}\left [ \ln  V (\xi^p) \right ]_{\xi^p} =
\lim_{n \to 0} \frac{\partial}{\partial n} \frac{1}{N}
\ln \left [  V^n (\xi^p) \right ]_{\xi^p} \cr
&& =\mathop{\rm Extr}_{\bTheta}
\left \{
\cA_0(\cw,\cu,\qw,\qu)+
\cA_w(\cw,\qw)+
\alpha \cA_u(\cu,\qu) 
\right \}, 
\label{replica_free_energy}
\end{eqnarray}
where $\bTheta=(\cw,\cu,\qw,\qu)$, 
\begin{eqnarray}
\cA_0(\cw,\cu,\qw,\qu)=F(\cw,\cu)+q_w \frac{\partial F(\cw,\cu)}{\partial \cw} 
-q_u \frac{\partial  F(\cw,\cu)}{\partial \cu}, 
\label{cA0}
\end{eqnarray}
\begin{eqnarray}
&&\cA_w(\cw,\qw)
=\mathop{\rm Extr}_{\hcw,\hqw}
\left \{
\frac{\hcw}{2}(\cw+\qw)-\frac{\hqw}{2}\cw \right .\cr
&& 
\left . \phantom{\cA_w}
+\int Dz \ln \left [\mathop{\rm Tr}_w P(w) 
\exp \left [
-\frac{\hcw}{2}w^2+\sqrt{\hqw}z w \right ]
\right ]
\right \}, 
\label{cAs}
\end{eqnarray}
and 
\begin{eqnarray}
&&\cA_u(\cu,\qu)
=\mathop{\rm Extr}_{\hcu,\hqu}
\left \{
\frac{\hcu}{2}(\cu-\qu)+\frac{\hqu}{2}\cu \right .\cr
&& 
\left . \phantom{\cA_w}
+\frac{1}{2}\mathop{\rm Tr}_y \int Dz 
\ln \left [
\int Dx \cI (y|\sqrt{\hcu} x+\sqrt{\hqu} z)
\right ]
\right \}. 
\label{cAu}
\end{eqnarray}
Here, $Ds=ds \exp \left [-s^2/2 \right ]/\sqrt{2 \pi}$ represents the Gaussian 
measure. 

Two points should be noted here. 
The first is that the current formalism can be applied not
only to the RS analysis presented above but also to that 
of replica symmetry breaking (RSB) \cite{MPV1987,Dotzenko2001}. 
An expression of the average free energy under the 
one-step RSB (1RSB) ansatz is shown in Appendix B. 
In addition, analysis of the local instability condition of 
the RS solution (\ref{Qs}) and (\ref{Qu})
subject to infinitesimal perturbation of the 
form of 1RSB yields
\begin{eqnarray}
\left (1-2 \frac{\partial^2 F}{\partial \cw^2}\cw^{(2)} \right )
\left (1-\frac{2}{\alpha} 
\frac{\partial^2 F}{\partial \cu^2}\cu^{(2)} \right )
-\frac{4}{\alpha}\left (\frac{\partial^2 F}{\partial \cw \partial \cu}
\right )^2 \cw^{(2)}\cu^{(2)} < 0, 
\label{AT}
\end{eqnarray}
where 
\begin{eqnarray}
\cw^{(2)}=\int Dz 
\left ( 
\frac{\partial^2}{\partial \left (\sqrt{\hqw}z\right )^2}
\ln \left [
\mathop{\rm Tr}_{w}
P(w) \exp \left [-\frac{\hcw}{2}w^2+\sqrt{\hqw}z w \right ]
\right ]
\right )^2, 
\label{cs2}
\end{eqnarray}
and 
\begin{eqnarray}
\cu^{(2)}=\frac{1}{2} \mathop{\rm Tr}_{y} \int Dz 
\left ( \frac{\partial^2}{\partial 
\left (\sqrt{\hqu}z\right )^2}
\ln 
\left [\int Dx \cI \left (y|\sqrt{\hcu}x+\sqrt{\hqu}z \right ) 
\right ]
\right )^2. 
\label{cu2}
\end{eqnarray}
Equation (\ref{AT}) corresponds to the de Almeida-Thouless (AT) condition 
for the current system \cite{AT1978}. 
The second point is that although randomly labeled patterns are
assumed here, one can develop a similar framework for 
analyzing the teacher-student scenario, which assigns pattern labels 
by a {\em teacher} perceptron, and 
which has a deep link to a certain class of 
modern wireless communication systems \cite{Tanaka2002,Kabashima2003,
Muller2003,Moustakas2003,GuoVerdu2005,NeirottiSaad2005,MontanariTse2006, 
MontanariPradhakarTse2005, TakedaUdaKabashima2006,TakedaHatabuKabashima2007}. 
One can find details of the framework
in reference \cite{Kabashima2007}.

\subsection{Thouless-Anderson-Palmer approach and message-passing algorithm}
The scheme developed so far is used for 
investigating typical macroscopic properties of perceptrons
which are averaged over pattern set ${\xi^p}$. 
However, another method is necessary to evaluate microscopic 
properties of a perceptron for an individual sample of $\xi^p$. 
The Thouless-Anderson-Palmer (TAP) approach \cite{TAP1978}, originating
in spin glass research, offers a useful guideline for this purpose. 
Although several formalisms are known for this approximation scheme 
\cite{OpperSaad2001}, we follow the one based on the Gibbs
free energy because of its generality and wide 
applicability \cite{OpperWinther2005,ParisiPotters1995}. 

Let us suppose a situation for which the microscopic averages 
of the dynamical variables,
\begin{eqnarray}
&&\bm_{w}=\mathop{\rm Tr}_{\bw} \bw P(\bw|\xi^p)  \cr
&&=\frac{1}{V(\xi^p)}
\mathop{\rm Tr}_{\bu,\bw} \bw \prod_{\mu=1}^p \widehat{\cI}_{y_\mu}(u_\mu)
\prod_{i=1}^N P(w_i) \exp \left [{\rm i} \bu^{\rm T}X \bw \right ], 
\label{micro_w_av}
\end{eqnarray}
and 
\begin{eqnarray}
\bm_{u}=\frac{1}{V(\xi^p)}
\mathop{\rm Tr}_{\bu,\bw} \left ({\rm i}\bu \right )
\prod_{\mu=1}^p \widehat{\cI}_{y_\mu}(u_\mu)
\prod_{i=1}^N P(w_i) \exp \left [{\rm i} \bu^{\rm T}X \bw \right ], 
\label{micro_u_av}
\end{eqnarray}
are required, where
$P(\bw|\xi^p)=\prod_{\mu=1}^p 
\cI(y_\mu|\Delta_\mu)\prod_{i=1}^N P(w_i)/V(\xi^p)$
denotes the posterior distribution of $\bw$ given $\xi^p$. 
The Gibbs free energy 
\begin{eqnarray}
\Phi(\bm_w,\bm_u)=
\mathop{\rm Extr}_{\bh_w,\bh_u}
\left \{ \bh_w \cdot \bm_w +\bh_u \cdot \bm_u -
\ln \left [
V(\bh_w,\bh_u)
\right]
\right \}, 
\label{gibbs_FE}
\end{eqnarray}
where 
\begin{eqnarray}
&&V(\bh_w,\bh_u) \cr
&&=\mathop{\rm Tr}_{\bu,\bw} 
\prod_{\mu=1}^p \widehat{\cI}_{y_\mu}(u_\mu)
\prod_{i=1}^N P(w_i) \exp \left [
\bh_w \cdot \bw+\bh_u \cdot ({\rm i}\bu)+
({\rm i} \bu)^{\rm T}X \bw \right ],
\label{h_partition}
\end{eqnarray}
offers a useful basis because the extremization 
conditions of equation (\ref{gibbs_FE}) generally 
agree with equations (\ref{micro_w_av}) and (\ref{micro_u_av}). 
This indicates that one can evaluate the microscopic averages
in equations (\ref{micro_w_av}) and (\ref{micro_u_av}) by extremization, 
which leads to assessment of the correct free energy, since
$\ln V(\xi^p)=-\mathop{\rm Extr}_{\{\bm_w,\bm_u\}}
\left \{ \Phi(\bm_w,\bm_u) \right \}$ holds, 
once the function of Gibbs free energy (\ref{gibbs_FE}) is provided. 

Unfortunately, an exact evaluation of equation (\ref{gibbs_FE}) is 
computationally difficult and therefore we resort to 
approximation. For this purpose, we put parameter $l$ 
in front of $X$ in equation 
(\ref{h_partition}), which yields the generalized Gibbs free energy as
\begin{eqnarray}
\widetilde{\Phi}(\bm_w,\bm_u;l)=\mathop{\rm Extr}_{\bh_w,\bh_u}
\left \{ \bh_w \cdot \bm_w +\bh_u \cdot \bm_u -
\ln \left [V(\bh_w,\bh_u;l)\right]
\right \}, 
\label{gen_gibbs_FE}
\end{eqnarray}
where $V(\bh_w,\bh_u;l)$
is defined by replacing $X$ with $lX$ in equation (\ref{h_partition}). 
This implies that the correct Gibbs free energy in equation (\ref{gibbs_FE}) can be 
obtained as $\Phi(\bm_w,\bm_u)=\widetilde{\Phi}(\bm_w,\bm_u;l=1)$
by setting $l=1$ in the generalized expression (\ref{gen_gibbs_FE}). 
One scheme for utilizing this relation is to perform 
the Taylor expansion around $l=0$, for which $\widetilde{\Phi}(\bm_w,\bm_u;l)$
can be analytically calculated as an exceptional case, and substitute 
$l=1$ in the expression obtained, which is sometimes 
referred to as the Plefka expansion \cite{Plefka1982}. 
However, evaluation of higher-order terms, 
which are non-negligible for correlated patterns in general, 
requires a complicated calculation in this expansion, 
which sometimes prevents the scheme from being practical. 
In order to avoid this difficulty, 
we take an alternative approach here, 
which is inspired by a derivative of equation (\ref{gen_gibbs_FE}), 
\begin{eqnarray}
\frac{\partial \widetilde{\Phi}(\bm_w,\bm_u;l) }{
\partial l}=-\left \langle ({\rm i} \bu)^T X \bw \right \rangle_l, 
\label{internal_energy}
\end{eqnarray}
where $\left \langle \cdots \right \rangle_l$ represents the 
average with respect to the generalized weight 
$\prod_{\mu=1}^p \widehat{\cI}_{y_\mu}(u_\mu)\times$
$\prod_{i=1}^N P(w_i)\times$ $\exp \left [
\bh_w \cdot \bw+\bh_u \cdot ({\rm i}\bu)+
({\rm i} \bu)^{\rm T}(lX) \bw \right ]$, and
$\bh_w$ and $\bh_u$ are determined to 
satisfy $\left \langle \bw \right \rangle_l=\bm_w$ and
$\left \langle ({\rm i}\bu) \right \rangle_l=\bm_u$, 
respectively \cite{OpperWinther2005}. 
The right-hand side of this equation is the 
average of a quadratic form containing many random variables. 
The central limit theorem 
implies that such an average does not depend on details 
of the objective distribution but is determined 
only by the values of the first and second moments. 
In order to construct a simple approximation scheme, 
let us assume that the second moments are
characterized macroscopically by 
$\left \langle |\bw|^2 \right \rangle_l-
|\left \langle \bw \right \rangle_l|^2 =N \cw$
and $\left \langle |\bu|^2 \right \rangle_l-
|\left \langle \bu \right \rangle_l|^2 =p \cu$. 
Evaluating the right-hand side of equation (\ref{internal_energy}) using 
a Gaussian distribution for which the first and 
second moments are constrained as 
$\left \langle \bw \right \rangle_l=\bm_w$, 
$\left \langle ({\rm i}\bu) \right \rangle_l=\bm_u$, 
$\left \langle |\bw|^2 \right \rangle_l-
|\left \langle \bw \right \rangle_l|^2 =N \cw$
and $\left \langle |\bu|^2 \right \rangle_l-
|\left \langle \bu \right \rangle_l|^2 =p \cu$, and 
integrating from $l=0$ to $l=1$ yields 
\begin{eqnarray}
&&\widetilde{\Phi}(\cw,\cu,\bm_w,\bm_u;1)
-\widetilde{\Phi}(\cw,\cu,\bm_w,\bm_u;0) \cr
&&\simeq -\bm_u^{\rm T} X \bm_w -N F(\cw,\cu), 
\label{AdaTAP}
\end{eqnarray}
where the function $F(x,y)$ is provided as in equation (\ref{F_func})
by the empirical eigenvalue 
spectrum of $X^{\rm T}X$, $\rho(\lambda)=N^{-1}
\sum_{k=1}^N \delta(\lambda-\lambda_k)$ and the 
macroscopic second moments $\cw$ and $\cu$ are included 
in arguments of the Gibbs free energy because the right-hand side 
of equation (\ref{internal_energy}) depends on them. 
Utilizing this and evaluating $\widetilde{\Phi}(\cw,\cu,\bm_w,\bm_u;0)$, which 
is not computationally difficult since interaction terms are not included, 
yield an approximation of the Gibbs free energy as
\begin{eqnarray}
&&\Phi(\cw,\cu,\bm_w,\bm_u) \simeq -\bm_u^{\rm T} X \bm_w -N F(\cw,\cu) 
\cr
&&+\mathop{\rm Extr}_{\hcw,\bh_w}
\left \{\bh_w \cdot \bm_w 
-\frac{1}{2}\hcw \left (N \cw+|\bm_w|^2 \right )
\right . \cr
&&\phantom{\mathop{\rm Extr}_{\hcw,\bh_w}
\left \{\bh_w \cdot \bm_w -\frac{1}{2}\hcw \right \}} 
 \left . -\sum_{i=1}^N \ln \left [\mathop{\rm Tr}_{w}
P(w) e^{-\frac{1}{2}\hcw w^2+h_{w i} w} \right ]
\right \} \cr
&&+\mathop{\rm Extr}_{\hcu,{\bh}_u}
\left \{\bh_u \cdot \bm_u
-\frac{1}{2}\hcu \left (p \cu-|\bm_u|^2 \right ) \right . \cr
&&\phantom{\mathop{\rm Extr}_{\hcw,\bh_w}
\left \{\bh_w \cdot \bm_w -\frac{1}{2}\hcw \right \}} 
\left . -\sum_{\mu=1}^p \ln \left [\int Dx
\cI(y_\mu|\sqrt{\hcu}x+h_{u \mu}) \right ]
\right \}, 
\label{TAP_free_energy}
\end{eqnarray}
which is a general expression of the {TAP free energy}
of the current system. Extremization of this equation 
provides a set of {TAP equations}
\begin{eqnarray}
m_{wi}&=&\frac{\partial}{\partial h_{wi}}
\ln \left [\mathop{\rm Tr}_{w}
P(w) e^{-\frac{1}{2}\hcw w^2+h_{w i} w} \right ], \label{TAPw1}\\
\cw&=&\frac{1}{N}\sum_{i=1}^N \frac{\partial^2}{\partial h_{wi}^2}
\ln \left [\mathop{\rm Tr}_{w} 
P(w) e^{-\frac{1}{2}\hcw w^2+h_{w i} w} \right ], \label{TAPw2}\\
m_{u\mu} &=& \frac{\partial}{\partial h_{u\mu}}
\ln \left [\int Dx
\cI(y_\mu|\sqrt{\hcu}x+h_{u \mu}) \right ], \label{TAPu1}\\
\cu &=& -\frac{1}{p}\sum_{\mu=1}^p \frac{\partial^2}{\partial h_{u\mu}^2}
\ln \left [\int Dx
\cI(y_\mu|\sqrt{\hcu}x+h_{u \mu}) \right ],\label{TAPu2}
\end{eqnarray}
where
\begin{eqnarray}
\bh_w&=&X^{\rm T} \bm_u -2\frac{\partial }{\partial \cw} F(\cw,\cu)
\bm_w, \label{TAPhw1} \\
\hcw&=&-2\frac{\partial }{\partial \cw} F(\cw,\cu), \label{TAPhw2} \\
\bh_u&=&X \bm_w 
+\frac{2}{\alpha}\frac{\partial }{\partial \cu} F(\cw,\cu)\bm_u, 
\label{TAPhu1} \\
\hcu&=&-\frac{2}{\alpha}\frac{\partial }{\partial \cu} F(\cw,\cu),  
\label{TAPhu2}
\end{eqnarray}
solutions of which represent approximate values of the first and 
second moments of the posterior distribution 
$P(\bw,\bu|\xi^p)$ for a fixed sample of $\xi^p$. 
In equations (\ref{TAPhw1}) and (\ref{TAPhu1}), 
$-2(\partial/\partial \cw) F(\cw,\cu) \bm_w$ and $ 
(2/\alpha) (\partial /\partial \cu )F(\cw,\cu)\bm_u$ 
are generally referred to as the {Onsager reaction terms}. 
The counterparts of these equations for systems of two-body 
interactions have been presented in an earlier paper 
\cite{ParisiPotters1995}. 

\begin{figure}
\small
\hspace*{1cm}{\bf {MPforPerceptron}}$\{$
\begin{eqnarray}
&&\mbox{Perform {\bf {Initialization}}}; \cr
&& \mbox{Iterate {\bf {H-Step}} and {\bf {V-Step}
} alternately sufficient times;}
\nonumber
\end{eqnarray}
\hspace*{1cm}$\}$\\
%\end{quote}
%\begin{quote}
\hspace*{1cm}{\bf {Initialization}}$\{$
\begin{eqnarray}
&&\cw \subst \frac{1}{N} \sum_{i=1}^N w_i^2 P(w_i); \quad \hcw \subst 0; 
\quad \Lambda_w \subst \frac{1}{\cw}-\hcw;\cr
&&m_{wi} \subst \mathop{\rm Tr}_{w_i}w_i P(w_i) \qquad (i=1,2,\ldots,N); \cr
&&\bh_u \subst X \bm_w;\quad \bm_h \subst \bzero;
\nonumber
\end{eqnarray}
\hspace*{1cm}$\}$\\
\hspace*{1cm}{\bf {H-Step}}$\{$
\begin{eqnarray}
&&\mbox{Search $(\cu,\Lambda_u)$ for given $(\cw,\Lambda_w)$ 
to satisfy conditions} \cr
&& \quad \cw=\left \langle \frac{\Lambda_u}{\Lambda_w \Lambda_u+\lambda} 
\right \rangle_\rho \mbox{ and } 
\cu=(1-\alpha^{-1})\frac{1}{\Lambda_u}
+\alpha^{-1} \left \langle \frac{\Lambda_w}{\Lambda_w \Lambda_u+\lambda} 
\right \rangle_\rho; \cr
&&\hcu\subst \frac{1}{\cu}-\Lambda_u; \cr
&& \bh_u \subst \bh_u-\hcu \bm_u; \cr
&& m_{u\mu} \subst \frac{\partial}{\partial h_{u\mu}}
\ln \left [\int Dx
\cI(y_\mu|\sqrt{\hcu}x+h_{u \mu}) \right ] \qquad (\mu=1,2,\ldots,p); \cr
&& \bh_w  \subst X^{\rm T} \bm_u; \cr
&& \cu \subst -\frac{1}{p}\sum_{\mu=1}^p \frac{\partial^2}{\partial h_{u\mu}^2}
\ln \left [\int Dx
\cI(y_\mu|\sqrt{\hcu}x+h_{u \mu}) \right ]; \cr
&& \Lambda_u \subst \frac{1}{\cu}-\hcu;
\nonumber
\end{eqnarray}
\hspace*{1cm}$\}$\\
\hspace*{1cm}{\bf {V-Step}}$\{$
\begin{eqnarray}
&&\mbox{Search $(\cw,\Lambda_w)$ for given $(\cu,\Lambda_u)$ 
to satisfy conditions} \cr
&& 
\quad \cw=\left \langle \frac{\Lambda_u}{\Lambda_w \Lambda_u+\lambda} 
\right \rangle_\rho \mbox{ and }
\cu=(1-\alpha^{-1})\frac{1}{\Lambda_u}
+\alpha^{-1} \left \langle \frac{\Lambda_w}{\Lambda_w \Lambda_u+\lambda} 
\right \rangle_\rho; \cr
&&\hcw\subst \frac{1}{\cw}-\Lambda_w; \cr
&& \bh_w \subst \bh_w+\hcw \bm_w; \cr
&& m_{w i} \subst \frac{\partial}{\partial h_{wi}}
\ln \left [\mathop{\rm Tr}_{w}
P(w) e^{-\frac{1}{2}\hcw w^2+h_{w i} w} \right ]
\quad (i=1,2,\ldots,N); \cr
&& \bh_u  \subst X \bm_w; \cr
&& \cw \subst 
\frac{1}{N}\sum_{i=1}^N \frac{\partial^2}{\partial h_{wi}^2}
\ln \left [\mathop{\rm Tr}_{w} 
P(w) e^{-\frac{1}{2}\hcw w^2+h_{w i} w} \right ]; \cr
&& \Lambda_w \subst \frac{1}{\cw}-\hcw;
\nonumber
\end{eqnarray}
\hspace*{1cm}$\}$
%\end{quote}
\normalsize
\caption{Pseudocode of the proposed message-passing algorithm 
{\bf MPforPerceptron}.
``;'' and ``$\leftarrow$'' represent 
the end of a command line and the operation of substitution, respectively. }
\label{fig1}
\end{figure}

Solving TAP equations (\ref{TAPw1})--(\ref{TAPhu2}) 
is not a trivial task. Empirically, naive iterative substitution of 
these equations does not converge in most cases. 
Conversely, it is reported that message-passing (MP) algorithms 
of a certain type, which are developed on the basis of 
the belief propagation \cite{Pearl1988},
exhibit excellent solution search performance for pattern 
sets entries of which are IID with low computational cost 
\cite{Kabashima2003,UdaKabashima2005}. 
Therefore, we developed an MP algorithm as a promising heuristic 
that reproduces known, efficient algorithms 
for IID pattern matrices. 
A pseudocode of the proposed algorithm is shown in figure \ref{fig1}. 
One can generalize this algorithm to the case of 
probabilistic perceptrons by replacing the indicator
function $\cI(y|\Delta)$ with a certain conditional 
probability $P(y|\Delta)$. 
It should be noted that $\Lambda_w$ and $\Lambda_u$ in the algorithm 
denote the counterparts of $\Lambda_x$ and $\Lambda_y$
in equation (\ref{F_func}) for $x=\cw$ and $y=\cu$, respectively. 
Solving $(\cu,\Lambda_u)$ and $(\cw,\Lambda_w)$
in {\bf H-Step} and {\bf V-Step}, respectively, 
can be performed efficiently by use of the bisection method. 
Solving the TAP equations employing this algorithm yields 
approximate estimates of the free energy $\ln V(\xi^p)$
and its derivatives as well as $\bm_w$ and $\bm_u$, 
which can be utilized for assessing whether the 
given specific sample $\xi^p$ can be correctly 
classified by the perceptron.

Although we have assumed single macroscopic 
constraints as characterizing the second moments, the current 
formalism can be generalized to include component-wise multiple constraints
for constructing more accurate approximations. 
By doing this, the current formalism leads to the adaptive TAP approach 
or, more generally, to the expectation consistent approximate schemes 
developed by Opper and Winther \cite{OpperWinther2001L,OpperWinther2001,
OpperWinther2005}. 

\section{Examples}
\subsection{Independently and identically distributed patterns}
In order to investigate the relationship with existing results,
let us first apply the developed methodologies to the case
in which the entries of $X$ are IID of zero mean and variance $N^{-1}$. 
This case can be characterized by the eigenvalue spectrum 
of the Mar\u{c}enko-Pastur type, which was already mentioned in section 2 and 
yields 
\begin{eqnarray}
F(x,y)=-\frac{\alpha }{2} xy. 
\label{IID_F_func}
\end{eqnarray}
This implies that equation (\ref{cA0}) can be expressed as
\begin{eqnarray}
\cA_0(\cw,\cu,\qw,\qu)
=-\frac{\alpha}{2} (\cw\cu + \qw\cu-\qu\cw).
\label{IID_A0}
\end{eqnarray}
Inserting this into equation (\ref{replica_free_energy}) 
and then performing an 
extremization with respect to $\cu$ and $\qu$ yields
\begin{eqnarray}
\hcu=\cw, \quad \hqu= \qw, 
\end{eqnarray}
where $\hcu$ and $\hqu$ are the variational variables 
used in equation (\ref{cAu}). 
This implies that the replica symmetric free energy (\ref{replica_free_energy})
can be expressed as
\begin{eqnarray}
&&\frac{1}{N}\left [\ln V(\xi^p) \right ]_{\xi^p} \cr
&& =\mathop{\rm Extr}_{\cw,\qw}\left \{
\cA_w(\cw,\qw) %\right .
%\left .
+ \frac{\alpha }{2} \mathop{\rm Tr}_y 
\int Dz \ln \left [
\int Dx \cI \left (y|\sqrt{\cw} x+\sqrt{\qw} z \right ) \right ] \right \}.  
\label{IID_replica_free_energy}
\end{eqnarray}
This is equivalent to the general expression of the replica symmetric 
free energy of a single-layer perceptron for the IID pattern matrices 
and randomly assigned labels \cite{Engel2001,OpperKinzel1993}.

\subsection{Rank deficient patterns vs. spherical weights}
In data analysis, the property of pattern components 
strongly correlated with each other
is referred to as {\em multicollinearity},  
which sometimes requires special treatment. 
As a second example, we utilize the developed framework to examine how this property influences 
$\alpha_c$. 

\begin{figure}[t]
\setlength{\unitlength}{1mm}
\begin{picture}(1870,65)
\put(10,0){\includegraphics[width=70mm]{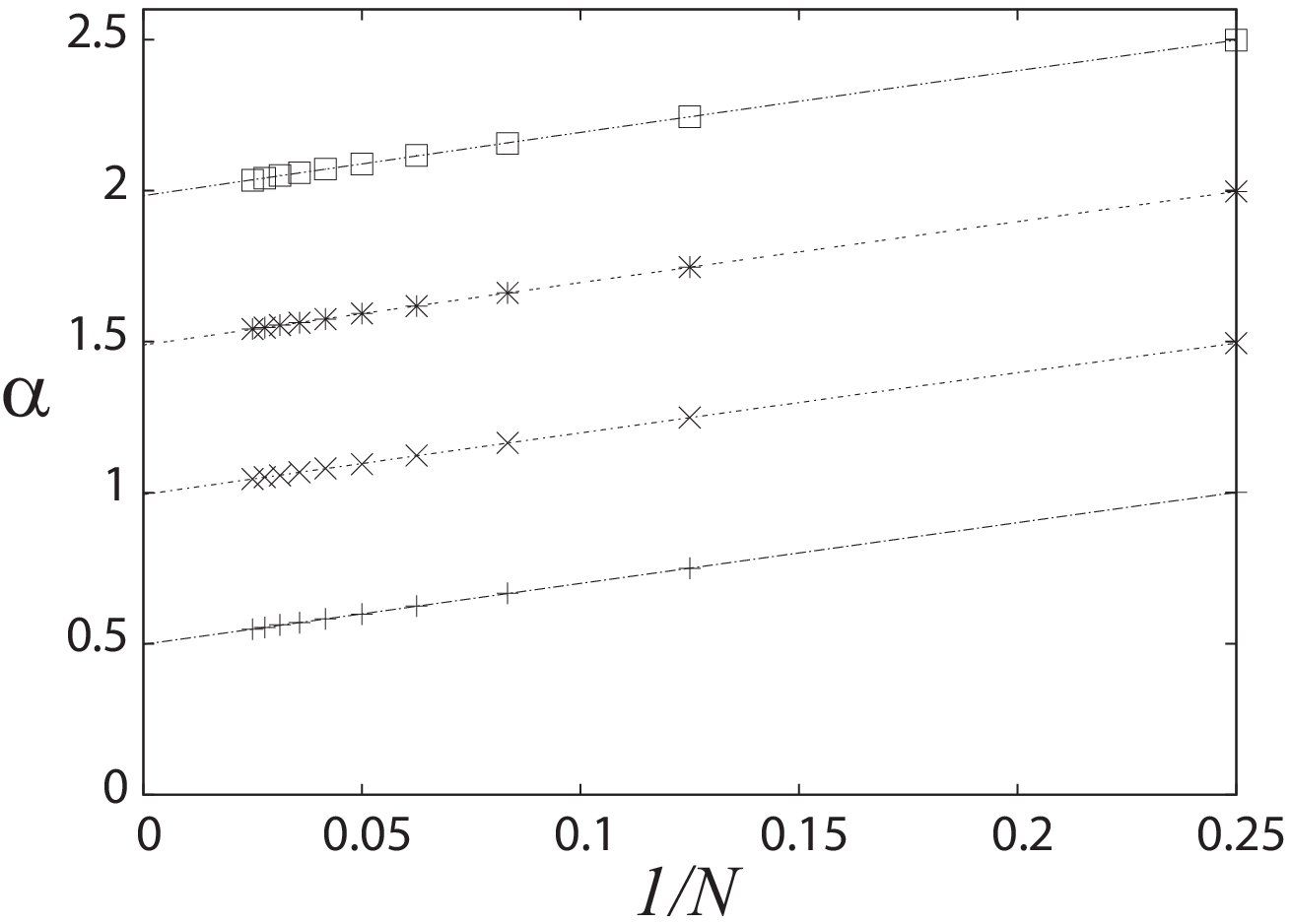}}
\put(85,0){\includegraphics[width=70mm]{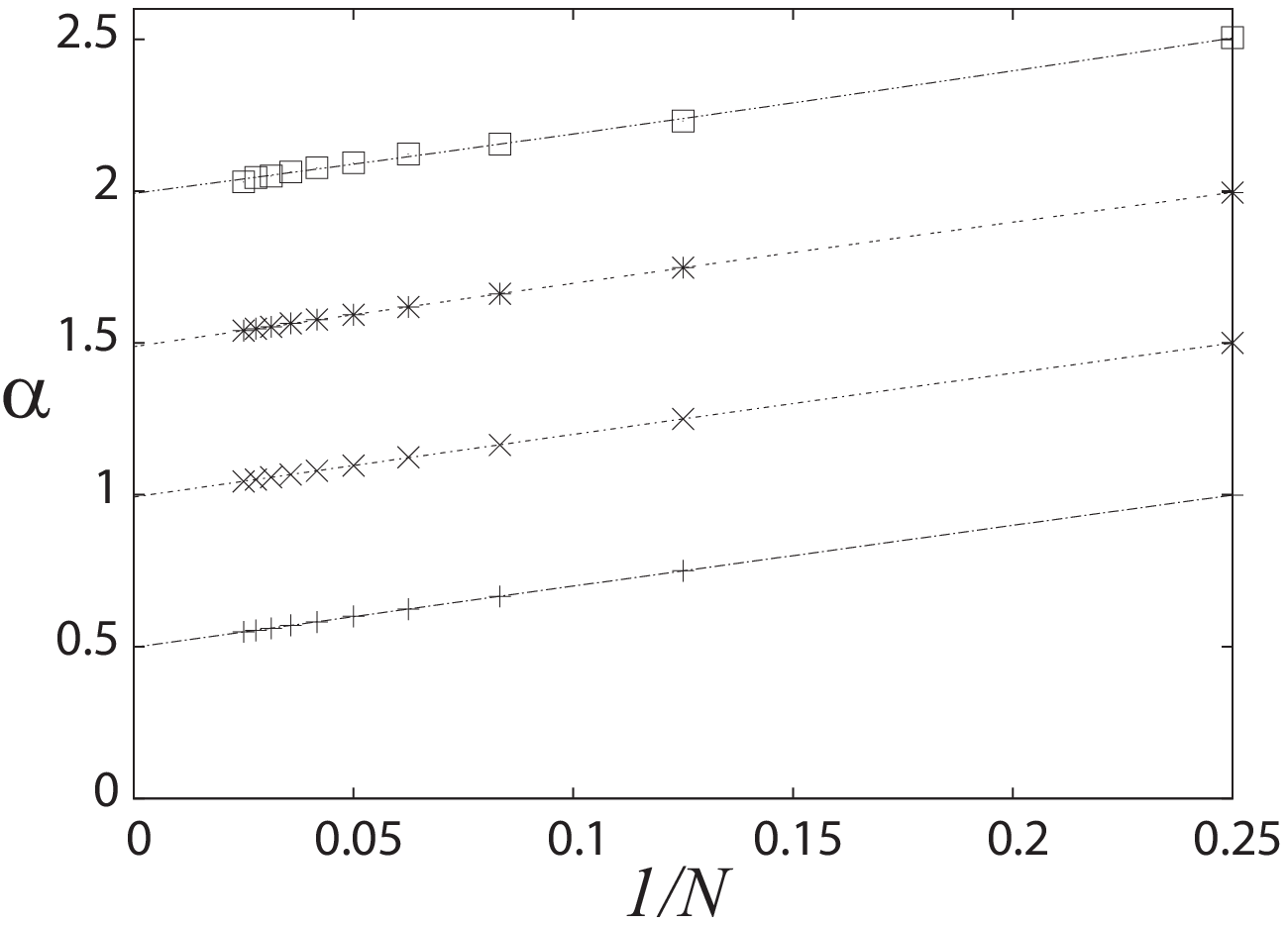}}
\put(15,55){(a)}
\put(90,55){(b)}
\end{picture}
\caption{
Assessment of $\alpha_c$ of spherical weights 
for rank deficient pattern matrices. For $N=4,8,12,\ldots,40$, 
the critical pattern ratio $\alpha_c(N)$, which is defined as 
the average of the maximum pattern ratio above which 
no weight can correctly classify a given sample of $\xi^p$, 
was assessed from $10^4$ experiments. 
Each estimate of $\alpha_c(N)$ was obtained by 
extrapolating $\alpha_c(N,T_{\rm max})$, which is 
an average value of $\alpha$ above which the perception 
learning algorithm \cite{MinskyPapaert1969}
does not converge after the number of updates reaches $T_{\rm max}$
for a given sample of $\xi^p$ with respect to $T_{\rm max}=10^3 
\sim 2 \times 10^4$. 
The capacity is estimated by a quadratic fitting 
under the assumption of $\alpha_c(N) \simeq \alpha_c + a N^{-1}+b N^{-2}$
where $a$ and $b$ are adjustable parameters. 
(a) and (b) represent results for 
$\widetilde{\rho}(\lambda)=
(2 \pi \lambda )^{-1}
\sqrt{[\lambda-(\sqrt{\alpha/c}-1)^2]^+ [(\sqrt{\alpha/c}+1)^2-
\lambda]^+}$ and 
$\delta(\lambda-1)$, respectively.
For both cases, each data corresponds to 
$c=1/4,1/2,3/4$ and $1$ from the bottom. 
The estimates of $\alpha_c$ show excellent consistency with 
the theoretical prediction $\alpha_c = 2 c$ regardless of 
$\widetilde{\rho}(\lambda)$. 
}
\label{fig2}
\end{figure}

Strong correlations among components can be modeled by 
rank deficiency of the cross-correlation 
matrix $X^{\rm T} X$. In the current 
framework, this is characterized by an eigenvalue
spectrum of the form 
\begin{eqnarray}
\rho(\lambda)=(1-c) \delta(\lambda)+c \widetilde{\rho}(\lambda), 
\label{rankdeficiency}
\end{eqnarray}
%for $\alpha>1$, 
where $0< c \le 1$ denotes the ratio between the rank of $X^{\rm T} X$ and $N$,
and $\tilde{\rho}(\lambda)$ is a certain distribution 
the support of which is defined over a region of $\lambda> 0$.
For simplicity, let us limit ourselves to the case
of simple perceptron and spherical weights, 
for which $\cI(y|\Delta)=1$ for $y\Delta >0$ and 
$0$, otherwise, and $P(\bw)=\delta(|\bw|^2-N)$. 
Inserting these into equation (\ref{replica_free_energy})
offers a set of saddle point equations.
Among them, those relevant for capacity analysis are 
\begin{eqnarray}
&&\cw=(1-c) \frac{1}{\Lambda_w}+
c\left \langle \frac{\Lambda_u}
{\Lambda_w \Lambda_u+\lambda} 
\right \rangle_{\widetilde{\rho}}, \label{sphSP1} \\
&&\cu=\left (1-\frac{c}{\alpha} \right ) \frac{1}{\Lambda_u}+
\frac{c}{\alpha} \left \langle \frac{\Lambda_w}{\Lambda_w 
\Lambda_u+\lambda} \right \rangle_{\widetilde{\rho}}, \label{sphSP2} \\
&& \hcu=-\frac{2}{ \alpha} \frac{\partial F(\cw,\cu)}{\partial \cu}
=\frac{1}{\cu}-\Lambda_u, \label{sphSP3} \\
&& \cu=-\int Dz \frac{\partial^2}{(\partial \sqrt{\widehat{q}_u}z )^2} 
\ln H\left (\frac{\sqrt{\widehat{q}_u}}{\sqrt{\hcu}}z 
\right ), \label{sphSP4}
\end{eqnarray}
where $H(x)=\int_x^{+\infty}Dz$. 

Let us assume that no RSB occurs for $\alpha < \alpha_c$, 
as is the case for IID patterns. Under this assumption, 
a critical condition is offered by taking a limit $\hcu \to 0$, 
which implies that the variance of $\bDelta
=(\Delta_1,\Delta_2,\ldots,\Delta_p)$ of 
the posterior distribution for a given sample $\xi^p$
typically vanishes. 
Applying an asymptotic form, $\ln H(x) \simeq -x^2/2$ for $x \gg 1$, 
to equation (\ref{sphSP4}) in conjunction with 
equation (\ref{sphSP3}) yields 
\begin{eqnarray}
\Lambda_u \simeq \frac{1}{2 \cu}. 
\label{lamda_u_capacity}
\end{eqnarray}
Inserting this into equation (\ref{sphSP2}) gives
\begin{eqnarray}
\frac{2c}{\alpha}-1 
\simeq \frac{c}{\alpha}
\left \langle \frac{2 \Lambda_w }{
\Lambda_w + 2\lambda \cu}
\right \rangle_{\widetilde{\rho}} \ge 0, 
\label{capacity_cond}
\end{eqnarray}
This means that no RS solution can exist for $\alpha > 2c $, 
indicating that the perceptron capacity is given as
\begin{eqnarray}
\alpha_c= 2c, 
\label{capacity2c}
\end{eqnarray}
regardless of $\widetilde{\rho}(\lambda)$. 
Equation (\ref{capacity2c}) is consistent with 
the known result $\alpha_c=2$ for IID patterns \cite{Cover1965,Gardner1988}, 
for which $c=1$ as $X^{\rm T}X$ is typically of full rank for $ \alpha > 1$. 
Numerical experiments for rank deficient pattern 
matrices support the present analysis, which is shown 
in figure \ref{fig2}. 
\begin{figure}[t]
\setlength{\unitlength}{1mm}
\begin{picture}(1870,65)
\put(10,0){\includegraphics[width=70mm]{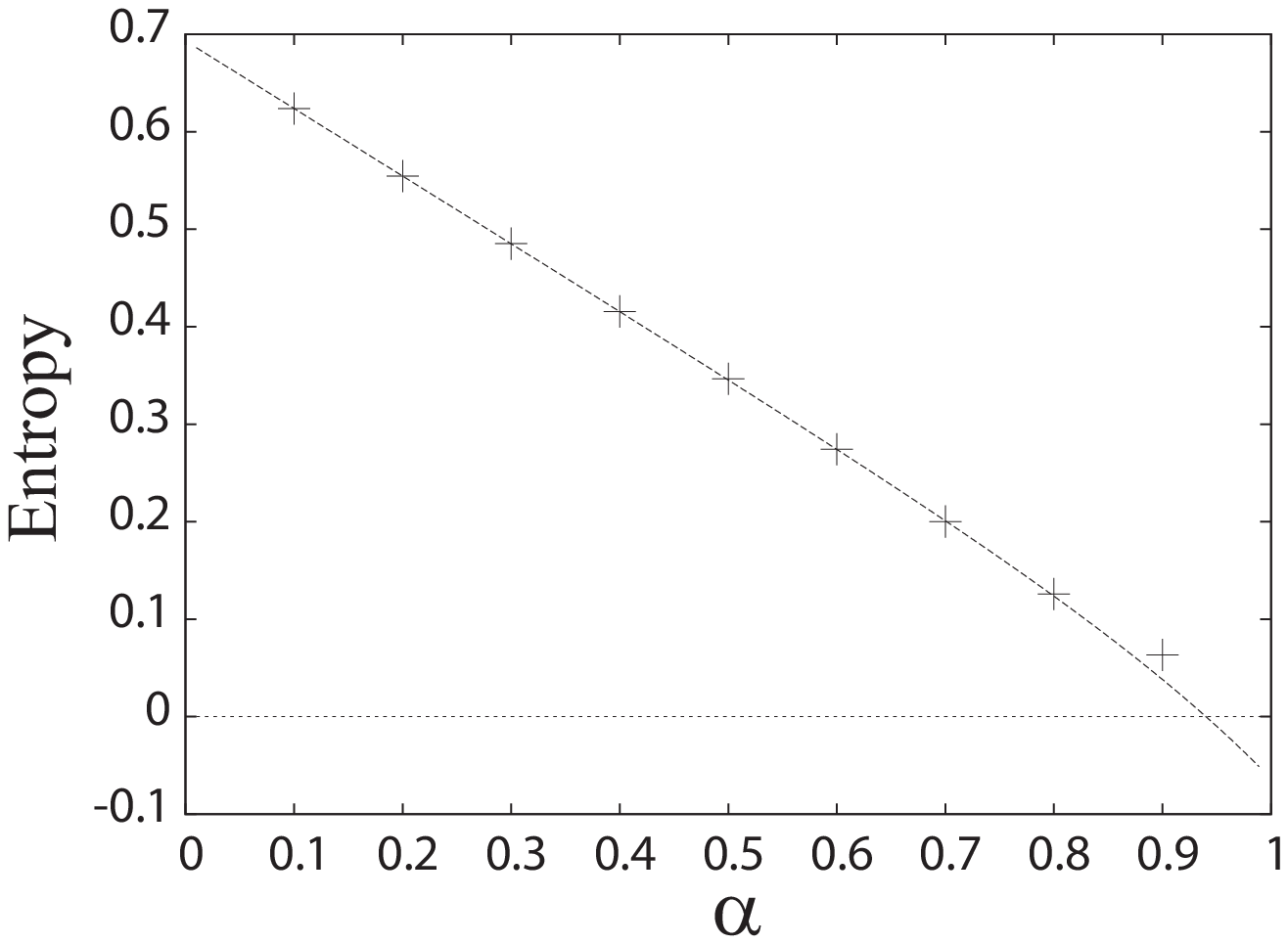}}
\put(85,0){\includegraphics[width=70mm]{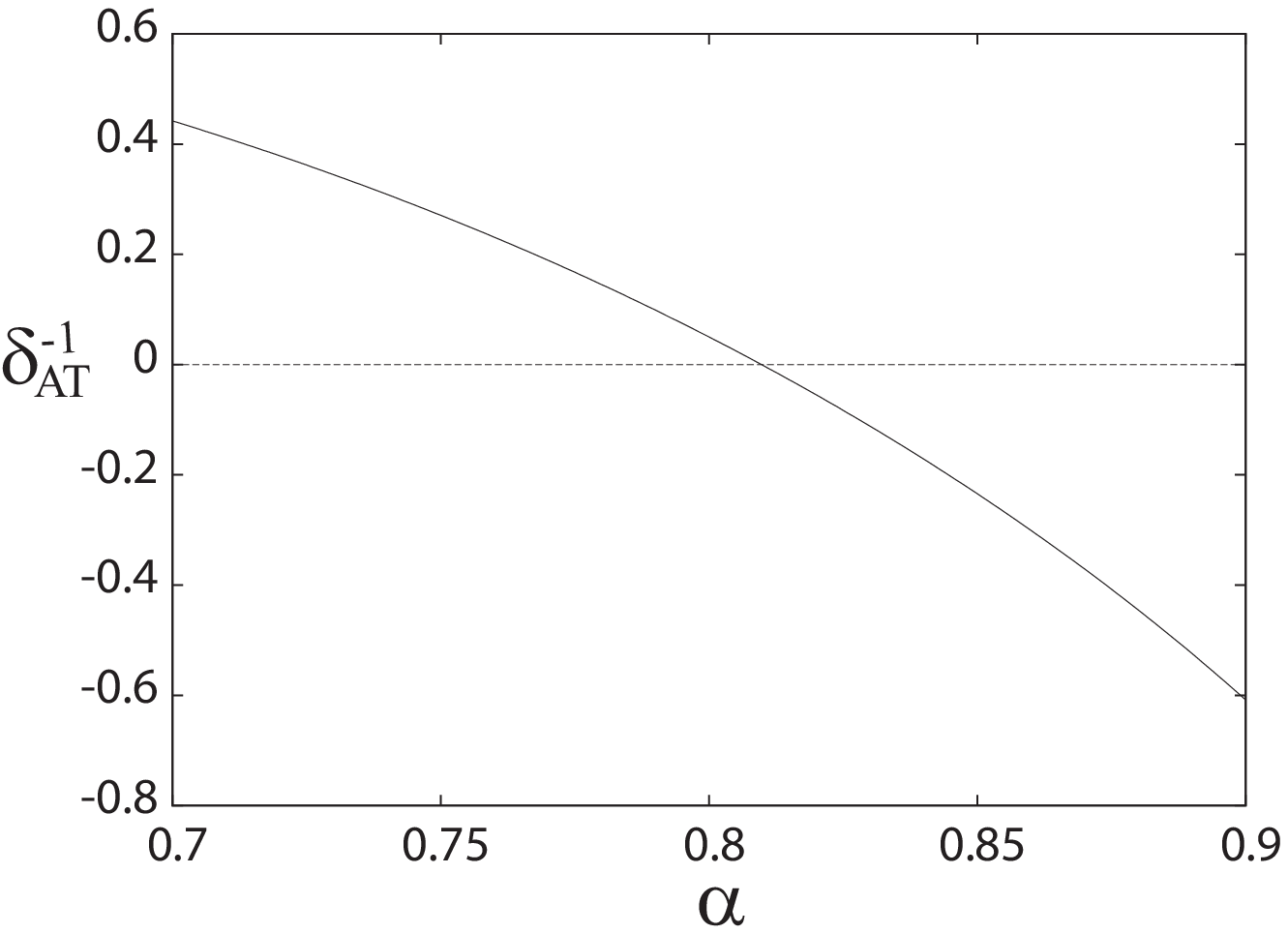}}
\put(15,55){(a)}
\put(90,55){(b)}
\end{picture}
\caption{(a): Entropy of $\bw$ (per element) versus the pattern ratio $\alpha$.
The curve represents the theoretical prediction assessed by the replica method
and the markers denote experimental data obtained by 
{\bf MPforPerceptron} for $100$ samples of $\xi^p$ of $N=500$ systems. 
(b): Diagnosis of the AT stability. $\delta_{\rm AT}^{-1}$, 
which is the inverse of the left-hand 
side of equation (\ref{AT}), is plotted versus $\alpha$
for the assessed RS solution. 
$\delta_{\rm AT}^{-1}$ becomes negative for 
$\alpha > \alpha_{\rm AT} \simeq 0.810$, indicating the
occurrence of RSB. 
}
\label{fig3}
\end{figure}

\begin{figure}[t]
\begin{center}
\setlength{\unitlength}{1mm}
\begin{picture}(1870,80)
\put(30,0){\includegraphics[width=100mm]{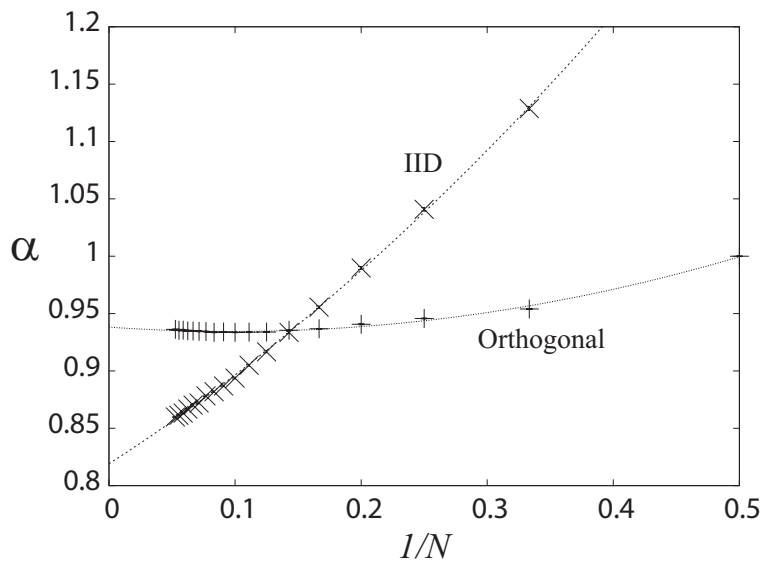}}
\end{picture}
\end{center}
\caption{Results of exhaustive search experiments. 
For $N=2,3,\ldots,20$, $\alpha_c(N)$, which is defined in the caption of 
figure \ref{fig2}, 
were estimated from $10^6$ experiments 
performed by an exhaustive search of binary weights. 
The values of capacity $\alpha_c$
are estimated by employing a quadratic fitting 
similar to that explained in 
the caption of figure \ref{fig2}. 
For IID patterns, this yields an estimate of $\alpha_c \simeq 0.819$, 
whereas the theoretical prediction is $0.833$ and is considered as
{exact}. The estimate $\alpha_c \simeq 0.938$ for the random 
orthogonal patterns is reasonably close to the theoretical 
prediction $0.940$, which is obtained from the unstable RS solution. }
\label{fig4}
\end{figure}

\subsection{Random orthogonal patterns vs. binary weights}
Equation (\ref{capacity2c}) means that 
the capacity depends only on the rank of 
the cross-correlation matrix $X^{\rm T}X$ 
in the case of spherical weights; 
however, this is not always the case. 
To show this, we present a capacity problem 
of binary weights $\bw=\{+1,-1\}^N$ as the final example. 

It is known that in typical cases, 
simple perceptrons of binary weights can correctly classify 
randomly labeled IID patterns 
for $\alpha < \alpha_c \simeq 0.833$
\cite{KrauthMezard1989,KrauthOpper1989,Derrida1991}. 
Our question here is how $\alpha_c$ is modified when 
the pattern matrix $X$ is generated randomly in such a way 
that patterns $\bx_\mu$ are orthogonal to each other. 

To answer this question, we employ the replica and 
TAP methods developed in the preceding sections 
for $\rho(\lambda)=(1-\alpha) \delta (\lambda)
+\alpha \delta(\lambda-1)$, which represents the eigenvalue spectrum 
of the random orthogonal patterns assuming $0< \alpha <1$ and 
yields
\begin{eqnarray}
F(x,y)=-1+\left (
\cL-\frac{1}{2} \ln \cL \right ), 
\label{Forth}
\end{eqnarray}
where $\cL=2^{-1}\left (1\pm \sqrt{1-4 \alpha xy} \right )$. 
Here, $\pm 1$ is chosen so that the operation 
of ${\rm Extr}_{\Lambda_x,\Lambda_y} \{ \cdots \}$ 
in equation (\ref{F_func})
corresponds to the correct saddle point evaluation of 
equation (\ref{bunshi}). 
Figure \ref{fig3} (a) shows how the entropy of $\bw$ depends on the 
pattern ratio $\alpha$. 
The curve denotes the theoretical prediction of the replica analysis
and the markers denote the averages of the entropy 
obtained by the TAP method over 100 samples 
for $N=500$ systems. The error bars are smaller than the markers. 
Solutions of the TAP method are obtained 
by {\bf MPforPerceptron}, shown in figure \ref{fig1}. 
Although the curve and the markers exhibit excellent agreement
for data points $\alpha =0.1,0.2,\ldots, 0.8$, 
we were not able to obtain a reliable result for $\alpha = 0.9$, 
at which point this algorithm does not converge in most cases, 
even after 1000 iterations.  
This may be a consequence of RSB since the replica analysis indicates that 
the AT stability of the RS solution shown in figure \ref{fig3} (a)
is broken for $\alpha$ beyond $\alpha_{\rm AT} \simeq 0.810$ 
(see figure \ref{fig3} (b)). 
Therefore $\alpha_c \simeq 0.940$, indicated by the condition 
of vanishing entropy is not regarded as the exact, but as an approximate 
value provided by the unstable RS solution. However, extrapolation 
of the results of direct numerical experiments 
for finite-size systems indicates that 
$\alpha_c \simeq 0.938$, as shown in figure \ref{fig4}, 
which implies that the effect of RSB is not significant 
for the evaluation of $\alpha_c$ in this particular case.

\section{Summary}
We developed a framework for analyzing 
the classification problems of perceptrons for 
randomly labeled patterns. The development is intended 
to handle correlated patterns. 
For this purpose, we developed two methodologies based on 
the replica method and the Thouless-Anderson-Palmer (TAP) approach, 
which are standard techniques from the statistical mechanics 
of disordered systems, and 
introduced a certain specific random assumption about the singular value 
decomposition of the pattern matrix. 
In both schemes, an integral formula, which 
can be regarded as a generalization of 
the Itzykson-Zuber integral known for square (symmetric)
matrices, plays an important role. 
As a promising heuristic for solving 
TAP equations. we provided
a message-passing algorithm {\bf MPforPerceptron}. 
The validity and utility of the developed 
schemes are shown for one known result and two novel problems. 

Investigation of the properties of {\bf MPforPerceptron}, as well as 
application of the developed framework to real-world 
data analysis \cite{UdaKabashima2005,Weigt2007}
and various models of information and communication engineering
\cite{Verdu1998,TulinoVerdu2004},
are promising topics for future research.

\ack
This research was supported in part by the JSPS Global COE program
``Computationism as a Foundation for the Sciences'' (TS and YK)
and Grants-in-Aid MEXT/JSPS, Japan, Nos. 17340116 and 18079006 (YK).

\appendix
\section{Derivation of Equations (\ref{generative_F}) and (\ref{F_func})}
The expressions 
\begin{eqnarray}
\delta(|\widetilde{\bw}|^2-Nx)=\frac{1}{2 \pi \img }
\int_{-\img \infty}^{+\img \infty}
 \frac{d \Lambda_x}{2} \exp \left [-\frac{\Lambda_x}{2}
(|\widetilde{\bw}|^2-Nx) \right ], \label{delta1} \\
\delta(|\widetilde{\bu}|^2-py)=\frac{1}{2 \pi \img }
\int_{-\img \infty}^{+\img \infty}
 \frac{d \Lambda_y}{2} \exp \left [-\frac{\Lambda_y}{2}
(|\widetilde{\bu}|^2-py) \right ], \label{delta2} 
\end{eqnarray}
yield an integral 
\begin{eqnarray}
&&\int d \widetilde{\bw} d\widetilde{\bu}
\delta(|\widetilde{\bw}|^2-Nx)
\delta(|\widetilde{\bu}|^2-py) 
\exp \left [\img \widetilde{\bu}^{\rm T} D \widetilde{\bw} 
\right ] \cr
&&=\frac{1}{(4 \pi \img )^2}
\int d \Lambda_x d \Lambda_y
\left (\int d \widetilde{\bw} d\widetilde{\bu} 
\exp \left [-\frac{\Lambda_x|\widetilde{\bw}|^2}{2}
-\frac{\Lambda_y|\widetilde{\bu}|^2}{2}+\img \widetilde{\bu}^{\rm T}D 
\widetilde{\bw} \right ] \right ) \cr
&& \phantom{\frac{1}{(4 \pi \img )^2}
\int d \Lambda_x d \Lambda_y}
\times \exp \left [\frac{N \Lambda_x x}{2}+\frac{p \Lambda_y y}{2} 
\right ]. \cr
&&=\frac{(2 \pi)^{(N+p)/2}}{(4 \pi \img )^2}
\int d \Lambda_x d \Lambda_y
\left (\det 
\left [
\begin{array}{ccc}
\Lambda_x I_N & \vline & -\img D^{\rm T} \cr
\hline 
-\img D & \vline & \Lambda_y I_p
\end{array}
\right ]
\right )^{-1/2} \cr
&&\phantom{\frac{1}{(4 \pi \img )^2}
\int d \Lambda_x d \Lambda_y}
\times \exp \left [\frac{N \Lambda_x x}{2}+\frac{p \Lambda_y y}{2} 
\right ], 
\label{bunshi}
\end{eqnarray}
where $I_N$ and $I_p$ are $N \times N$ and $p \times p$ identity
matrices, respectively. Linear algebra can be used to generate the expression 
\begin{eqnarray}
&&\ln \det 
\left [
\begin{array}{ccc}
\Lambda_x I_N & \vline & -\img D^{\rm T} \cr
\hline 
-\img D & \vline & \Lambda_y I_p
\end{array}
\right ]
=\sum_{k=1}^{{\rm min}(p,N)} \ln (\Lambda_x \Lambda_y+\lambda_k)\cr
&& \phantom{\ln \det 
\left [
\begin{array}{ccc}
\Lambda_x I_N & \vline & \img D^{\rm T} \cr
\hline 
\img D & \vline & \Lambda_y I_p
\end{array}
\right ]=N+}
+ (N-{\rm min}(p,N))\ln \Lambda_y  \cr
&& \simeq N 
\left (
\left \langle \ln (\Lambda_x \Lambda_y +\lambda ) \right \rangle_\rho 
+(\alpha -1)\ln \Lambda_y \right ), 
\end{eqnarray}
in the large system limit
$N,p \to \infty$, keeping $\alpha=p/N \sim O(1)$. 
This implies that equation (\ref{bunshi}) can be evaluated by the 
saddle point method as 
\begin{eqnarray}
&&\frac{1}{N}\ln \left [
\int d \widetilde{\bw} d\widetilde{\bu}
\delta(|\widetilde{\bw}|^2-Nx)
\delta(|\widetilde{\bu}|^2-py) 
\exp \left [\img \widetilde{\bu}^{\rm T} D \widetilde{\bw} 
\right ] \right ] \cr
&& =
\mathop{\rm Extr}_{\Lambda_x,\Lambda_y}\!
\left \{\! -\!
\frac{1}{2}\left \langle \ln (\Lambda_x \Lambda_y+\lambda)
\right \rangle_\rho 
\!- \! \frac{\alpha-1}{2}\ln \Lambda_y+\frac{\Lambda_x x}{2}
\! +\! \frac{\alpha \Lambda_y y}{2}\right \} \!+\! const, 
\label{logbunshi}
\end{eqnarray}
where $const$ represents constant terms that do not 
depend on either $x$ or $y$.  
In particular, setting $D=0$ in this expression leads to
\begin{eqnarray}
&&\frac{1}{N}\ln \left [
\int d \widetilde{\bw} d\widetilde{\bu}
\delta(|\widetilde{\bw}|^2-Nx)
\delta(|\widetilde{\bu}|^2-py) 
\right] \cr
&& =\mathop{\rm Extr}_{\Lambda_x,\Lambda_y}
\left \{- \frac{1}{2}\ln \Lambda_x 
-\frac{\alpha}{2}\ln \Lambda_y
+\frac{\Lambda_x x}{2}+\frac{\alpha \Lambda_y y}{2} \right \} \cr
&&=\frac{1}{2}\ln x +\frac{\alpha}{2}\ln y +\frac{1+\alpha}{2}
+ const. 
\label{logbunbo}
\end{eqnarray}
Equations (\ref{logbunshi}) and (\ref{logbunbo})
are used in equations (\ref{generative_F}) and (\ref{F_func}). 

\section{Assessment of free energy under the 1RSB ansatz}
The argument in section 3 implies that when $n \times n$ matrices 
$\cQ_w=(q_w^{ab})$ and $\cQ_u=(q_u^{ab})$ are simultaneously 
diagonalized by an identical orthogonal matrix, 
the average of the replicated coupling term 
with respect to $U$ and $V$ is evaluated as
\begin{eqnarray}
\frac{1}{N}\ln 
\left [\overline{\exp \left [
\img \sum_{a=1}^n (\bu^a)^{\rm T} X \bw^a \right ] }
\right ]=\sum_{a=1}^n F(t_w^a,t_u^a), 
\label{Fdecomposition}
\end{eqnarray}
where $t_w^a$ and $t_u^a$ ($a=1,2,\ldots,n$) denote a pair 
of eigenvalues of $\cQ_w$ and $\cQ_u$
and correspond to an identical eigen vector.
Under the 1RSB ansatz, $n$ replica indices are divided into 
$n/m$ groups of identical size $m$, and the relevant saddle 
point is characterized as
\begin{eqnarray}
(q_w^{ab},q_u^{ab})=\left \{
\begin{array}{ll}
(\cw+v_w+q_w,\cu-v_u-q_u), & a=b, \cr
(v_w + q_w, -v_u-q_u), & \mbox{$a$ and $b$ belong} \cr
& \mbox{to an identical group},\cr
(q_w,-q_u), &\mbox{otherwise}, 
\end{array}
\right .
\label{1RSBansatz}
\end{eqnarray}
where $m$ serves as Parisi's RSB parameter after analytical continuation. 
$\cQ_w$ and $\cQ_u$ of the form of equation (\ref{1RSBansatz}) can be 
simultaneously diagonalized, which yields pairs of eigenvalues as
\begin{eqnarray}
(t_w^a,t_u^a)=
\left \{
\begin{array}{ll}
(\cw  +  m v_w+ n q_w, \cu - m v_u -n q_u), & 1, \cr
(\cw+m v_w, \cu- m v_u), & n/m-1, \cr
(\cw, \cu), & n-n/m, 
\end{array}
\right .
\end{eqnarray}
where the numbers in the right-most column represent the degeneracies 
of the pair of eigenvalues denoted in the middle column. 
This gives 
\begin{eqnarray}
&&\frac{1}{N}\ln 
\left [\overline{\exp \left [
\img \sum_{a=1}^n (\bu^a)^{\rm T} X \bw^a \right ] }
\right ] \cr
&&= F(\cw \!+ \!m v_w \!+ \!n q_w, \cu \!-\! m v_u \!- \!n q_u) +
\left (\frac{n}{m}\! -\! 1 \right )F(\cw\!+ \! m v_w, \cu \!- \! m v_u) \cr
&&\phantom{=}+\left (n-\frac{n}{m} \right )F(\cw,\cu). 
\label{1RSB_Fxy}
\end{eqnarray}
Equation (\ref{1RSB_Fxy}) and assessment of 
the volumes of dynamical variables $\{\bw^a\}$ and $\{\bu^a\}$
under the 1RSB ansatz (\ref{1RSBansatz}), 
in conjunction with analytical continuation 
from $n \in \mN$ to $n \in \mR$, lead to the 
expression of the 1RSB free energy as
\begin{eqnarray}
&& \frac{1}{N}\left [ \ln V(\xi^p) \right ]_{\xi^p} =\lim_{n \to 0}
\frac{\partial }{\partial n} 
\frac{1}{N}\ln \left [ V^n(\xi^p) \right ]_{\xi^p} \cr
&&=\mathop{\rm Extr}_{\bTheta,m} 
\left \{ \cA_0^{\rm 1RSB}(\cw,\cu,v_w,v_u,q_w,q_u;m) \right .\cr
&&
\left . \phantom{\mathop{\rm Extr}_{\bTheta} 
\left \{ \cA_0^{\rm 1RSB} \right \}}
+\cA_w^{\rm 1RSB}(\cw,v_w,q_w;m)
+\alpha \cA_u^{\rm 1RSB}(\cu,v_u,q_u;m) \right \}, 
\label{1RSBfree_energy}
\end{eqnarray}
where $\bTheta=(\cw,\cu,v_w,v_u,q_w,q_u)$, 
\begin{eqnarray}
&&\cA_0^{\rm 1RSB}(\cw,\cu,v_w,v_u,q_w,q_u;m) \cr
&&=F(\cw,\cu)+\frac{1}{m} \left (F(\cw+m v_w,\cu-m v_u)-F(\cw,\cu) \right ) \cr
&& 
\phantom{==}+ q_w \frac{\partial F(\cw+m v_w,\cu-m v_u)}{\partial \cw}-
q_u \frac{\partial F(\cw+m v_w,\cu-m v_u)}{\partial \cu}, 
\label{1RSBfree_energyA0}
\end{eqnarray}
\begin{eqnarray}
&&\cA_w^{\rm 1RSB}(\cw,v_w,q_w;m) \cr
&&=\mathop{\rm Extr}_{\hcw,\widehat{v}_w,\hqw}
\left \{\frac{\hcw (\cw+v_w+q_w)}{2}-
\frac{\widehat{v}_w(\cw+m(v_w+q_w))}{2}-\frac{\hqw(\cw+m v_w)}{2} \right . \cr
&&\phantom{==}
\left . + \frac{1}{m}\int Dz \ln \left [
\int Dy \left (
\mathop{\rm Tr}_{w} P(w)e^{-\frac{\hcw}{2}w^2+(\sqrt{\widehat{v}_w}y
+\sqrt{\hqw} z )w} \right )^m \right ] \right \}, 
\label{1RSBfree_energyAw}
\end{eqnarray}
and 
\begin{eqnarray}
&&\cA_u^{\rm 1RSB}(\cu,v_u,q_u;m) \cr
&&=\mathop{\rm Extr}_{\hcu,\widehat{v}_u,\hqu}
\left \{\frac{\hcu (\cu-v_u-q_w)}{2}+
\frac{\widehat{v}_u (\cu-m(v_u+q_u))}{2}+\frac{\hqu(\cu-m v_u)}{2} \right . \cr
&&\phantom{==}
\left . + \frac{1}{2 m} \mathop{\rm Tr}_y \!
\int \! Dz \! \ln \left [\! \int \! Ds \left (
\! \int \! Dx \! \cI(y|\sqrt{\hcu}x+\sqrt{\widehat{v}_u} s \!+ \!
\sqrt{\hqu} z )
\! \right  )^m \! \right ] \! \right \}. 
\label{1RSBfree_energyAu}
\end{eqnarray}

\end{document}